\documentclass{sty/jpaper}
\input{sty/mysty}

\newif\ifcameraready
\camerareadytrue

\newcommand{\versionnum}[0]{4.1}

\ifcameraready
\else
\fi

\ifcameraready
  \newcommand{\todo}[1][]{}
  \newcommand{\ch}[0]{}
\else
  \newcommand{\todo}[1][]{\textbf{\fcolorbox{black}{red}{\color{white}{TODO}}} \underline{$\overline{\hbox{\emph{#1}}}$}}
  \newcommand{\ch}[1]{{\color{BrickRed} #1}}
\fi

\title{Tiered-Latency DRAM:\\ Enabling \ch{Low-Latency Main Memory} at Low Cost}

\author{
  Donghyuk Lee$^{1,2}$\qquad
  Yoongu Kim$^{2}$\qquad
  Vivek Seshadri$^{3,2}$
  \vspace{2pt}\\
  Jamie Liu$^{4,2}$\qquad
	Lavanya Subramanian$^{5,2}$\qquad
  Onur Mutlu$^{6,2}$}
\affil{
  $^1${\em NVIDIA Research}\qquad
  $^2${\em Carnegie Mellon University}\vspace{2pt}\\
  $^3${\em Microsoft Research India}\qquad
  $^4${\em Google}\qquad
  $^5${\em Intel Labs}\qquad
  $^6${\em ETH Z{\"u}rich}\\
}

\begin{document}
	\date{}
	\maketitle

	\begin{abstract}

This paper summarizes the idea of Tiered-Latency DRAM (TL-DRAM), 
which was published in HPCA 2013~\cite{lee-hpca2013}, and examines
the work's significance and future potential.
The capacity and cost-per-bit of DRAM have historically scaled to satisfy the
needs of increasingly large and complex computer systems. However, DRAM latency
has remained almost constant, making memory latency the performance bottleneck
in today's systems. We observe that the high access latency is not intrinsic to
DRAM, but a trade-off is made to decrease the cost per bit. To mitigate the high area
overhead of DRAM sensing structures, commodity DRAMs connect many DRAM cells to
each sense amplifier through a wire called a bitline. These bitlines have a high
parasitic capacitance due to their long length, and this bitline capacitance is
the dominant source of DRAM latency. Specialized low-latency DRAMs use shorter
bitlines with fewer cells, but have a higher cost-per-bit due to greater
sense amplifier area overhead.

To achieve both low latency and low cost per bit, we introduce {\em Tiered-Latency DRAM} (TL-DRAM).
In TL-DRAM, each long bitline is split
into two shorter segments by an isolation transistor, allowing one of the two segments to be
accessed with the latency of a short-bitline DRAM without incurring a high
cost per bit. We propose mechanisms that use the low-latency segment as a
hardware-managed or software-managed cache. Our evaluations show that our proposed
mechanisms improve both performance and energy efficiency for both single-core
and multiprogrammed workloads.

\ch{Tiered-Latency DRAM has} inspired several other works
on reducing DRAM latency with little to no architectural modification~\cite{lee-hpca2015, 
chang-hpca2014, chang-sigmetrics2016, chang-sigmetrics2017, lee-sigmetrics2017,
hassan-hpca2016, seshadri-micro2013, seshadri-micro2015, seshadri-micro2017,
chang-thesis2017, hassan-hpca2017}.

\end{abstract}

	\section{Problem: High DRAM Latency} \label{sec:problem}

Primarily due to its low cost per bit, DRAM has long been the substrate of choice
for architecting main memory subsystems. In fact, DRAM's cost per bit has been
decreasing at a rapid rate as DRAM process technology scales to integrate ever
more DRAM cells into the same die area. As a result, each successive generation
of DRAM has enabled increasingly larger-capacity main memory subsystems at low
cost.

\begin{figure}[ht]
\centering
\vspace{2pt}
\includegraphics[width=1\linewidth]{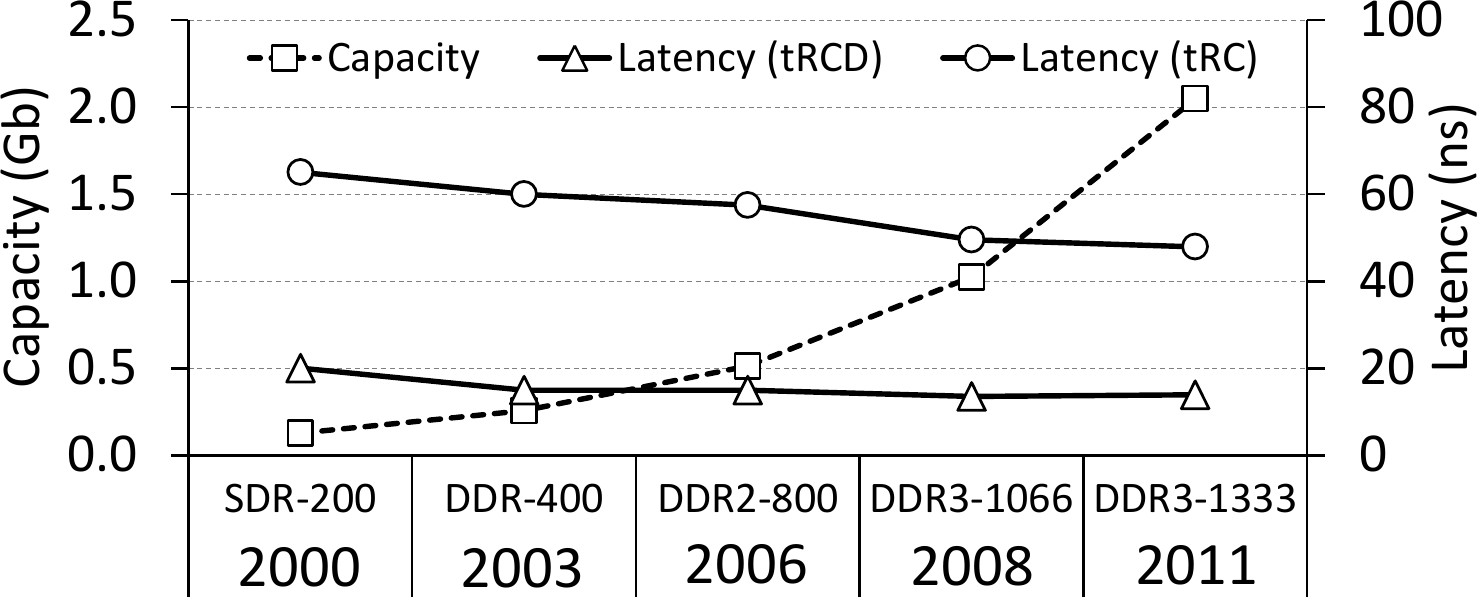}
\vspace{2pt}
\caption{Change in DRAM capacity and latency over time~\cite{borkar-cacm2011, jung-2005,
patterson-2004, samsung_spec}. Reproduced from \cite{lee-hpca2013}.}
\label{fig:intro_trend}
\end{figure}

In stark contrast to the continued scaling of cost per bit, the {\em latency} of
DRAM has remained almost constant. During the same 11-year interval in which
DRAM's cost per bit decreased by a factor of 16, DRAM latency (as measured by
the \trcd and \trc timing constraints)\footnote{The overall DRAM latency can be
decomposed into individual DRAM {\em timing constraints}. Two of the most
important timing constraints are \trcd (row-to-column delay) and \trc (row-cycle
time).} decreased by only 30.5\% and 26.3\%~\cite{borkar-cacm2011, jung-2005}, respectively,
as shown in Figure~\ref{fig:intro_trend}. From the perspective of the processor,
an access to DRAM takes hundreds of cycles -- time during which the processor
may be stalled, waiting for DRAM~\cite{mutlu-hpca2003, mutlu-isca2005,
mutlu-ieeemicro2003, ailamaki-vldb1999, kanev-isca2015, ghose-isca2013}. 
This wasted time due to stalling on DRAM leads to large performance degradation.

\section{Key Observations and Our Goal} \label{sec:observation}

{\bf Bitline: Dominant Source of Latency.} In DRAM, each bit is represented as
electrical charge in a capacitor-based {\em cell}. The small size of this
capacitor necessitates the use of an auxiliary structure, called a {\em sense
amplifier}, to (1)~detect the small amount of charge held by the cell and (2)~amplify it
to a full digital logic value. A sense amplifier is approximately one hundred
times larger than a cell~\cite{rambus-power}. To amortize their large size, each
sense amplifier is connected to many DRAM cells through a wire called a {\em
bitline}.\footnote{We refer the reader to our prior works for a detailed background
on DRAM architecture and operation~\cite{kim-isca2012, kim-micro2010,
kim-hpca2010, lee-hpca2013, lee-hpca2015, chang-hpca2014, chang-hpca2016,
chang-sigmetrics2016, chang-sigmetrics2017, hassan-hpca2016, hassan-hpca2017,
liu-isca2012,
liu-isca2013, patel-isca2017, qureshi-dsn2015, seshadri-micro2013,
seshadri-micro2017, lee-sigmetrics2017, kim-cal2015, lee-pact2015, lee-taco2016,
kim-isca2014, kim-hpca2018}.} 

Every bitline has an associated \emph{parasitic capacitance}, whose value is
proportional to the length of the bitline. Unfortunately, the parasitic
capacitance slows down DRAM operation for two reasons. First, it increases the
latency of the sense amplifiers. When the parasitic capacitance is large, a cell
cannot quickly create a voltage perturbation on the bitline that can be easily
detected by the sense amplifier. Second, the capacitance increases the latency
of charging and precharging the bitlines. Although the cell and the bitline must
be restored to their quiescent voltages during and after an access to a cell,
such a procedure takes much longer when the parasitic capacitance of the bitline is large. Due
to these two reasons, and based on a detailed latency breakdown discussed in
Section~3.1 of our HPCA 2013 paper~\cite{lee-hpca2013}, we conclude that long
bitlines are the dominant source of DRAM latency~\cite{jedec-ddr, dram_latency,
mutlu-imw13, mutlu-book15, lee-hpca2013, lee-hpca2015}.

\begin{figure}[ht]
  \centering
  \begin{subfigure}[b]{1in}
    \centering
    \captionsetup{font=footnotesize}
    \includegraphics[width=0.9in]{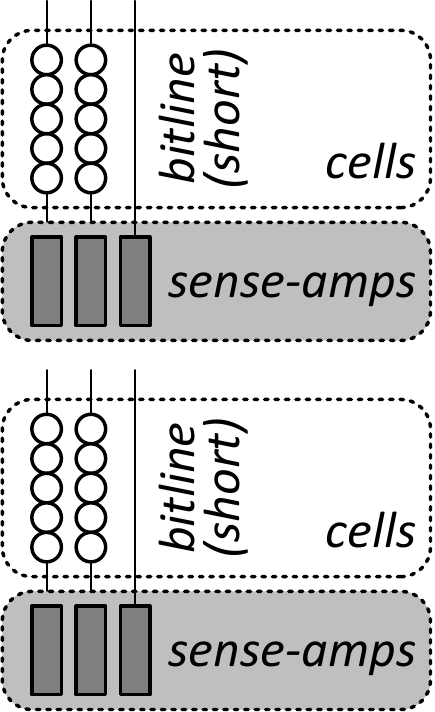}
    \subcaption{Latency-optimized architecture}
    \label{fig:intro_specialized_dram}
  \end{subfigure}~~
  \begin{subfigure}[b]{1.1in}
    \centering
    \captionsetup{font=footnotesize}
    \includegraphics[width=0.9in]{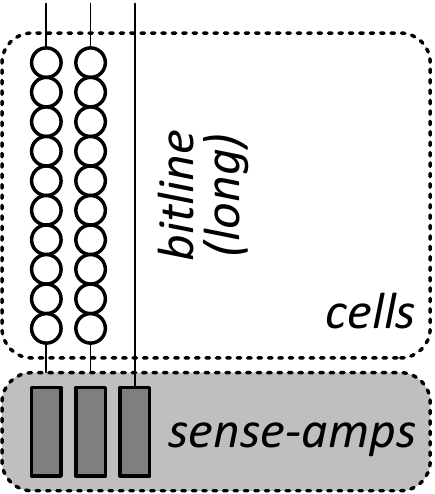}
    \subcaption{Cost-optimized architecture}
    \label{fig:intro_commodity_dram}
  \end{subfigure}~~
  \begin{subfigure}[b]{1in}
    \centering
    \captionsetup{font=footnotesize}
    \includegraphics[width=0.9in]{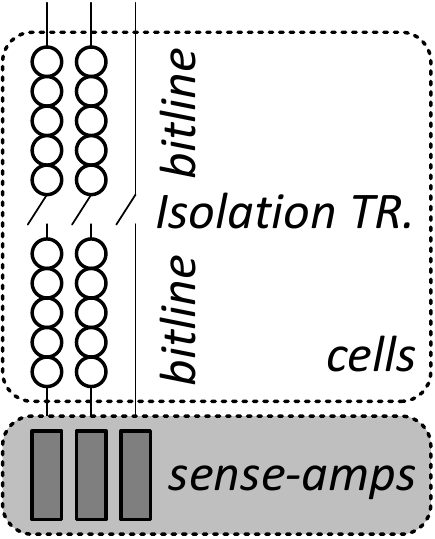}
    \subcaption{Our proposed architecture}
    \label{fig:intro_tldram}
  \end{subfigure}
  \vspace{5pt}
  \caption{DRAM latency and cost optimization, and our proposal (TL-DRAM). Reproduced from~\cite{lee-hpca2013}.}
  \label{fig:intro_commodity_specialized}
\end{figure}

{\bf Latency vs.~Cost Trade-Off.} The bitline length is a key design parameter
that exposes the important trade-off between latency and die size (cost). Short
bitlines (i.e., a bitline connected to only a few cells) constitute a small
electrical load (parasitic capacitance), which leads to low latency. However,
they require more sense amplifiers for a given DRAM capacity
(Figure~\ref{fig:intro_specialized_dram}), which leads to a large die size. In
contrast, long bitlines have high latency and a small die size
(Figure~\ref{fig:intro_commodity_dram}). As a result, neither of these two
approaches can optimize for {\em both} latency and cost per bit.

\begin{figure}[h]
	\vspace{0.1in}
  \centering
  \includegraphics[width=\linewidth]{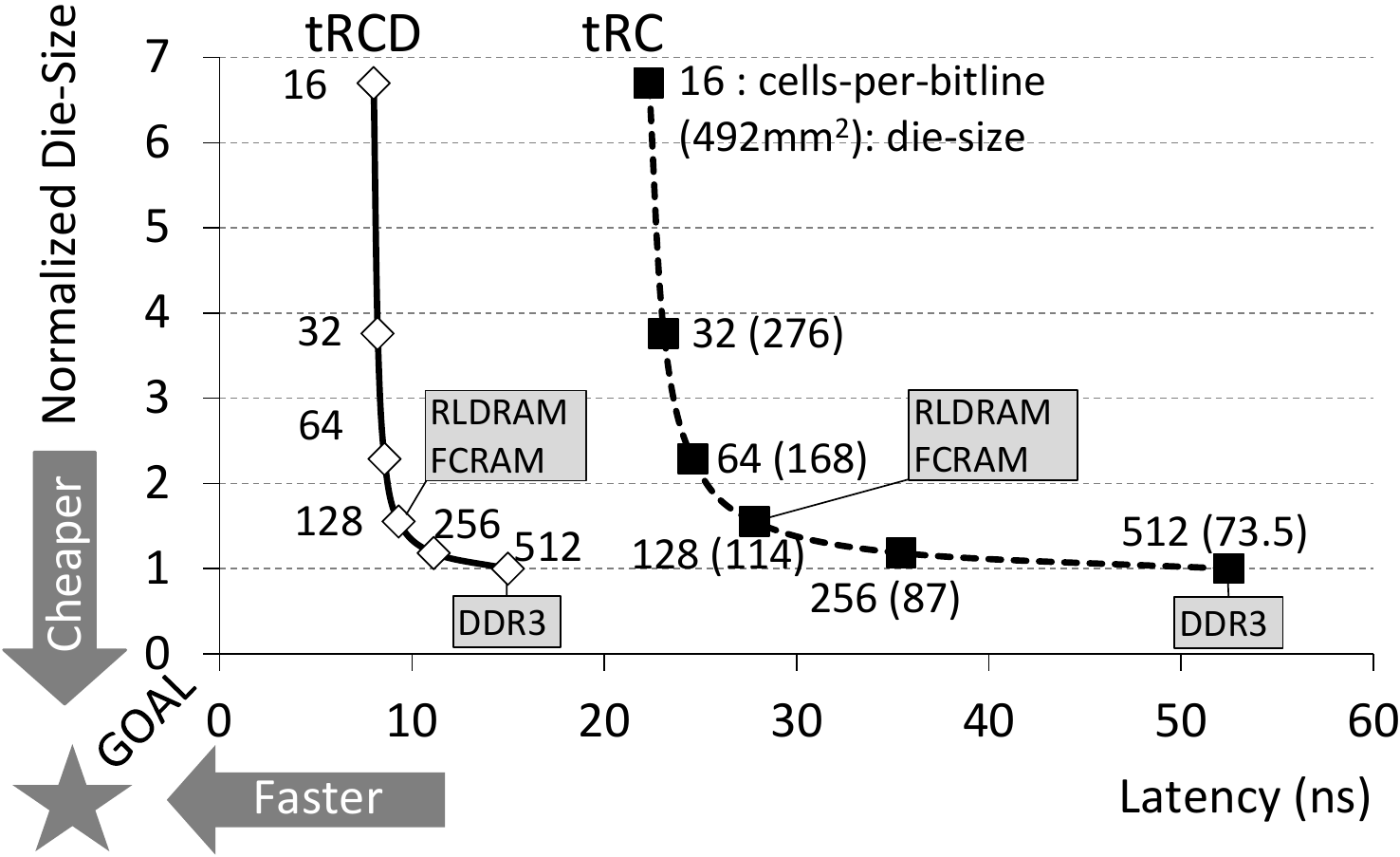}
  \caption{Bitline length: latency vs. die size. Reproduced from~\cite{lee-hpca2013}.}
  \label{fig:cell-per-bitline-trade-off}
\end{figure}

Figure~\ref{fig:cell-per-bitline-trade-off} shows the trade-off between DRAM
latency and die size by plotting the latency (\trcd and \trc) and the die size
for different values of cells per bitline. Existing DRAM architectures are
either (1)~optimized for die size (e.g., commodity DDR3 \cite{samsung_spec, ddr3-4gb}) and
are thus low cost but high latency; or (2)~optimized for latency (e.g.,
RLDRAM~\cite{rldram}, FCRAM~\cite{fcram}) and are thus low latency but (very)
high cost.

{\bf The goal} of our HPCA 2013 paper~\cite{lee-hpca2013} is to design a new
DRAM architecture to approximate the best of both worlds (i.e., low latency \emph{and}
low cost), based on our key observation that long bitlines are the dominant
source of DRAM latency.

\section{Tiered-Latency DRAM} \label{sec:tldram}

\begin{figure}[ht]
  \begin{subfigure}[b]{0.9in}
    \centering
    \includegraphics[height=1.2in]{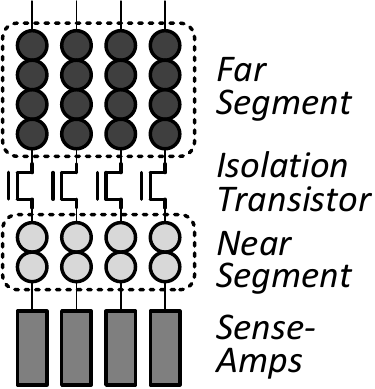}
    \subcaption{Organization}
    \label{fig:substrate_tld}
  \end{subfigure}
	\hspace{0.20in}
  \begin{subfigure}[b]{1.1in}
    \centering
    \includegraphics[height=1.2in]{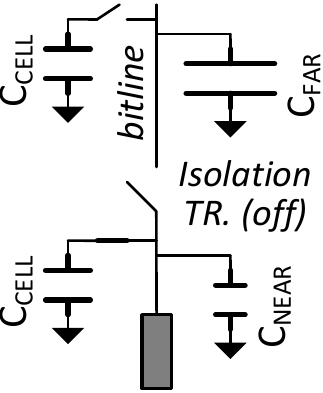}
    \caption{Near segment}
    \label{fig:substrate_tld_near}
  \end{subfigure}
	\hspace{-0.05in}
  \begin{subfigure}[b]{1.1in}
    \centering
    \includegraphics[height=1.2in]{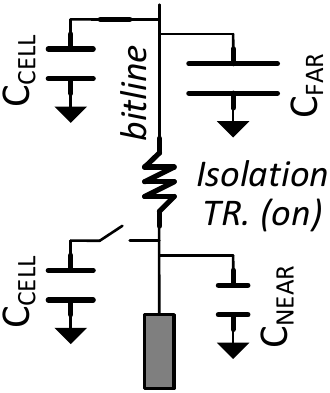}
    \caption{Far segment}
    \label{fig:substrate_tld_far}
  \end{subfigure}
	\caption{TL-DRAM: accessing the near segment and the far segment. 
	Adapted from~\cite{lee-hpca2013}.}
  \label{fig:substrate_tldram}
\end{figure}

To achieve the latency advantage of short bitlines {\em and} the cost advantage
of long bitlines, we propose the {\em Tiered-Latency DRAM} (TL-DRAM)
architecture, which is shown in Figures~\ref{fig:intro_tldram} and
\ref{fig:substrate_tld}. The key idea of TL-DRAM is to divide the long bitline
into two shorter segments using an {\em isolation transistor}: the {\em near
segment} (connected directly to the sense amplifier) and the {\em far segment}
(connected through the isolation transistor).

The primary role of the isolation transistor is to electrically decouple the two
segments from each other. This changes the effective bitline length (and also
the effective bitline capacitance) as seen by the cell and sense amplifier.
Correspondingly, the latency to access a cell also changes, albeit differently
depending on whether the cell is in the near or the far segment.

When accessing a cell in the near segment, the isolation transistor is turned
off, disconnecting the far segment (Figure~\ref{fig:substrate_tld_near}). Since
the cell and the sense amplifier see only the reduced bitline capacitance of the
shortened near segment, they can drive the bitline voltage more easily. As a
result, the bitline voltage is restored more quickly, and, thus, the latency
(\trc) for the near segment is significantly reduced. On the other hand, when
accessing a cell in the far segment, the isolation transistor is turned on to
connect the entire length of the bitline to the sense amplifier. In this case,
the isolation transistor acts like a resistor inserted between the two segments
(Figure~\ref{fig:substrate_tld_far}) and limits how quickly charge flows to the
far segment. Because the far segment capacitance is charged more slowly, it
takes longer for the far segment voltage to be restored, and, thus, the latency
(\trc) is increased for cells in the far segment.

\begin{figure*}[t]

\begin{minipage}[b]{0.31\linewidth}
  \centering
  \begin{subfigure}[b]{\linewidth}
    \centering
    \includegraphics[width=\linewidth]{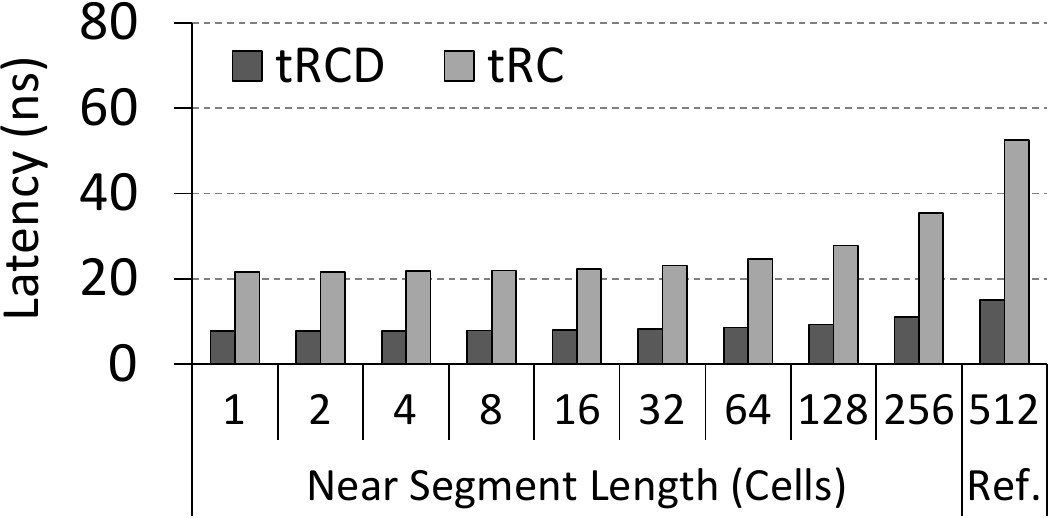} \\\vspace{-0.08in}
    \subcaption{Cell in near segment}
    \label{fig:substrate_latency_near}
  \end{subfigure}\\ \vspace{0.1in}
  \begin{subfigure}[b]{\linewidth}
    \centering
    \includegraphics[width=\linewidth]{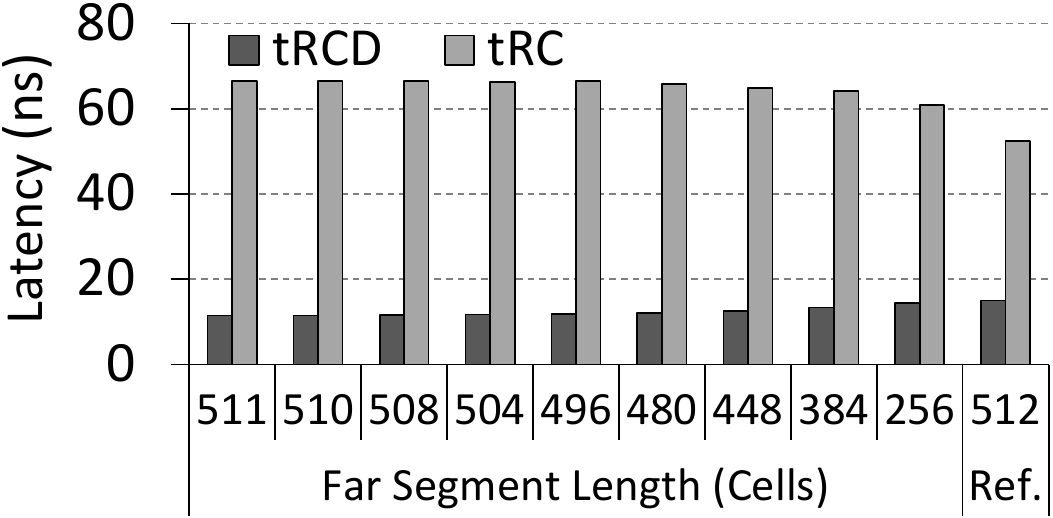} \\\vspace{-0.08in}
    \subcaption{Cell in far segment}
    \label{fig:substrate_latency_far}
  \end{subfigure}
  \centering
	\caption{Latency analysis. Reproduced from~\cite{lee-hpca2013}.} \label{fig:substrate_latency}
\end{minipage}\qquad
\begin{minipage}[b] {0.39\linewidth}
  \centering
  \begin{subfigure}[b]{\linewidth}
  \centering
	\hspace{-15mm} \footnotesize{\trcdnear} \hspace{1.7mm} \footnotesize{\trasnear} \\
    \includegraphics[width=\linewidth]{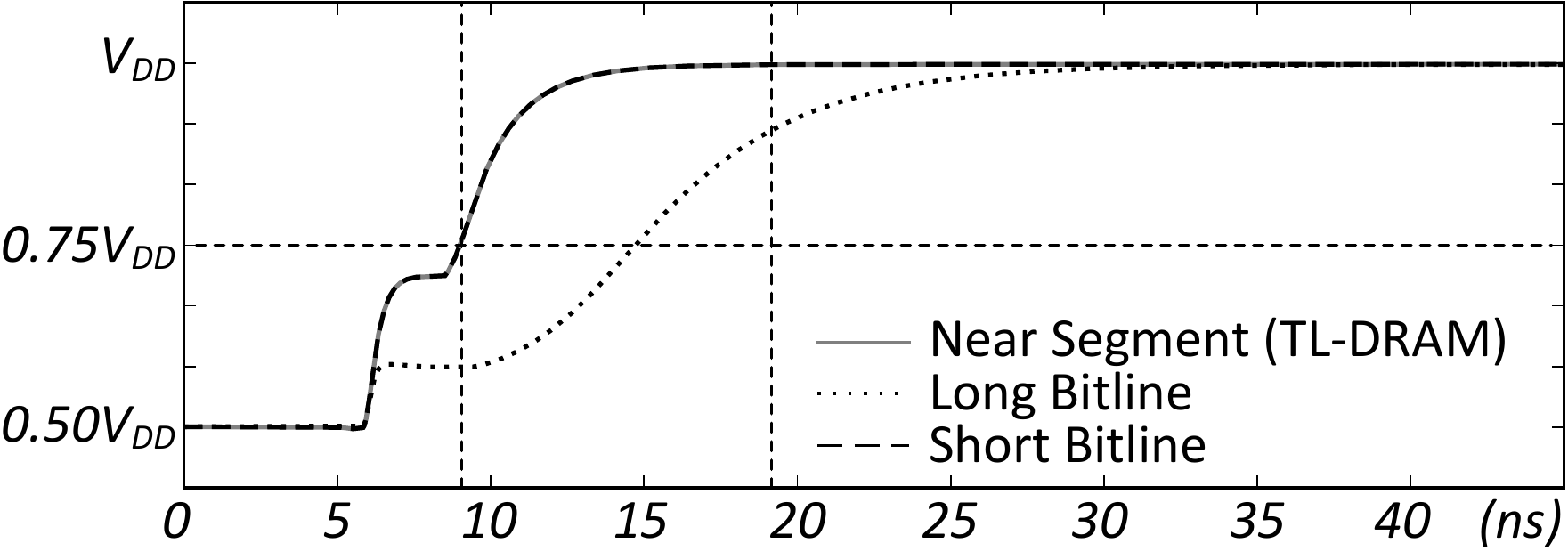} \\ \vspace{-0.08in}
    \subcaption{Cell in near segment (128 cells)}
    \label{fig:substrate_sim_active_off}
  \end{subfigure}\\ \vspace{0.1in}
  \begin{subfigure}[b]{\linewidth}
  \centering
	\hspace{20mm} \footnotesize{\trcdfar} \hspace{30mm} \footnotesize{\trasfar} \\
    \includegraphics[width=\linewidth]{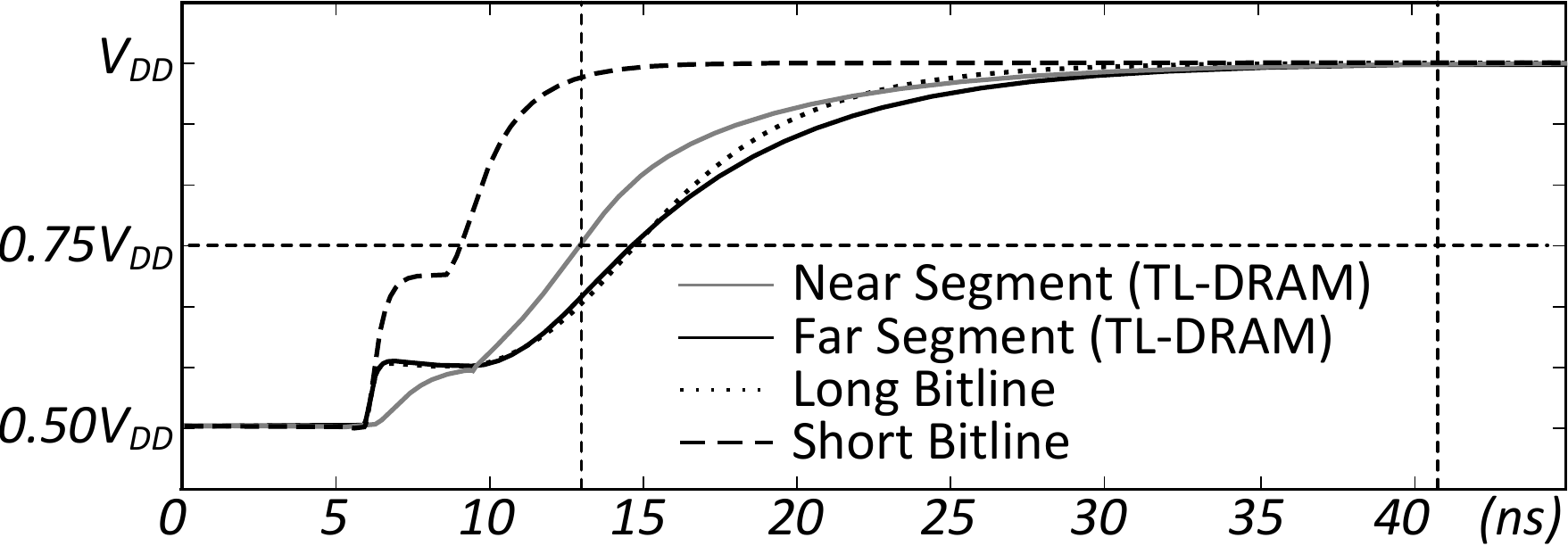} \\\vspace{-0.08in}
    \subcaption{Cell in far segment (384 cells)}
    \label{fig:substrate_sim_active_on}
  \end{subfigure}
  \caption{Activation: bitline voltage. Reproduced from~\cite{lee-hpca2013}.} \label{fig:substrate_sim_active}
\end{minipage}\qquad
\begin{minipage}[b] {0.21\linewidth}
  \centering
  \begin{subfigure}[b]{\linewidth}
  \centering
	\hspace{-2mm} \footnotesize{\trpnear}
    \includegraphics[width=\linewidth]{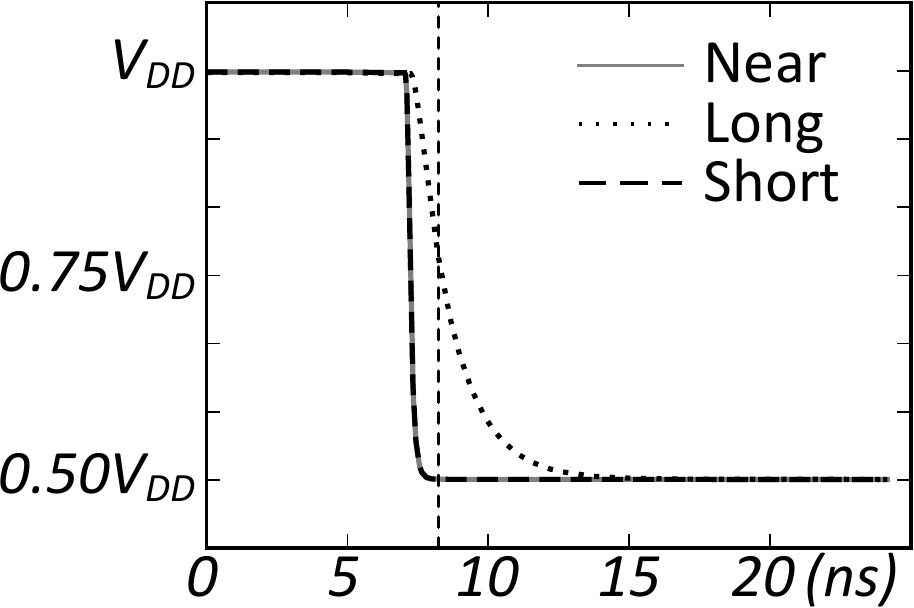} \\\vspace {-0.08in}
    \subcaption{Cell in near segment}
    \label{fig:substrate_sim_prech_off}
  \end{subfigure}\\ \vspace{0.1in}
  \begin{subfigure}[b]{\linewidth}
  \centering
	\hspace{28mm} \footnotesize{\trpfar} 
    \includegraphics[width=\linewidth]{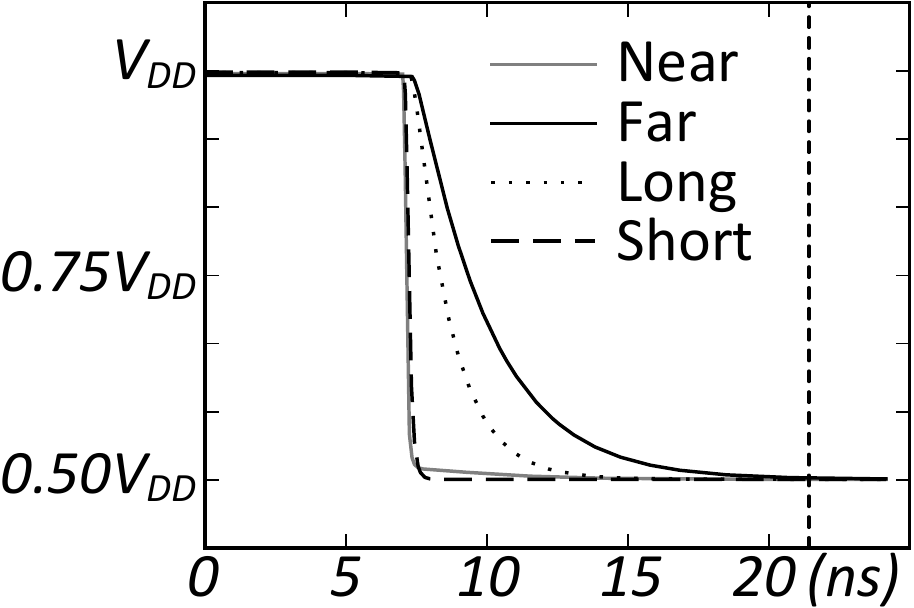} \\\vspace {-0.08in}
    \subcaption{Cell in far segment}
    \label{fig:substrate_sim_prech_on}
  \end{subfigure}
  \centering
  \caption{Precharging. Reproduced from~\cite{lee-hpca2013}.} \label{fig:substrate_sim_prech}
\end{minipage}

\end{figure*}

{\bf Sensitivity to Segment Length.} The lengths of the two segments are
determined by where the isolation transistor is placed on the bitline. Assuming
that the number of cells per bitline is fixed at 512 cells, the near segment
length can range from as short as a single cell to as long as 511 cells. We
perform circuit-level simulations to determine how the latency of each segment
based on the number of cell in the segment.
Figures~\ref{fig:substrate_latency_near} and~\ref{fig:substrate_latency_far}
plot the latencies of the near and far segments as a function of their length,
respectively. For reference, the rightmost bars in each figure are the latencies
of an unsegmented long bitline whose length is 512 cells. From these figures, we
draw three conclusions. First, the shorter the near segment, the lower its
latencies (\trcd and \trc). This is expected since a shorter near segment has a
lower effective bitline capacitance, allowing it to be driven to target voltages
more quickly. Second, the longer the far segment, the lower the far segment's
\trcd. Recall from our previous discussion that the far segment's \trcd depends
on how quickly the {\em near segment} (not the far segment) can be driven. A
longer far segment implies a shorter near segment (lower capacitance), which is
why \trcd decreases for the far segment. Third, the shorter the far segment, the
smaller its \trc. The far segment's \trc is determined by how quickly it reaches
the full voltage (\vdd or {\em 0}). Regardless of the length of the far segment
or the near segment, the current that trickles into it through the isolation
transistor does not change significantly. Therefore, a shorter far segment
(lower capacitance) reaches the full voltage more quickly.

{\bf Latency Analysis (Circuit Evaluation).} We model TL-DRAM in detail using
SPICE simulations. Simulation parameters are mostly derived from a publicly
available 55nm DDR3 2Gb process technology file~\cite{rambus-power} which
includes information such as cell and bitline capacitances and resistances,
physical floorplanning, and transistor dimensions. Transistor device
characteristics were derived from~\cite{ibm_55nm} and scaled to agree
with~\cite{rambus-power}. Figures~\ref{fig:substrate_sim_active}
and~\ref{fig:substrate_sim_prech} show the bitline voltages during activation
and precharging, respectively. The $x$-axis origin (time 0) in the two figures
corresponds to when the subarray receives the \cmdact or \cmdpre command,
respectively. In addition to the voltages of the segmented bitline (near and far
segments), the figures also show the voltages of two unsegmented bitlines (short
and long) for reference.

First, during an access to a cell in the near segment
(Figure~\ref{fig:substrate_sim_active_off}), the far segment is disconnected and
is floating (hence its voltage is not shown). The bitline starts at $1/2$ \vdd.
Due to the reduced bitline capacitance of the near segment, its voltage
increases almost as quickly as the voltage of a short bitline (the two curves
are overlapped) during {\em sensing \& amplification}. Since the near segment
voltage reaches 0.75\vdd and \vdd (the {\em threshold} and {\em restored}
states, respectively) quickly, its \trcd and \tras, respectively, are
significantly reduced compared to a long bitline. Second, during an access to a
cell in the far segment (Figure~\ref{fig:substrate_sim_active_on}), we can
indeed verify that the voltages of the near and the far segments increase at
different rates due to the resistance of the isolation transistor, as previously
explained. Compared to a long bitline, while the near segment voltage reaches
0.75\vdd more quickly, the far segment voltage reaches \vdd more slowly. As a
result, \trcd for the far segment is reduced while its \tras is increased.

While precharging the bitline after accessing a cell in the near segment
(Figure~\ref{fig:substrate_sim_prech_off}), the near segment reaches 0.5\vdd
quickly due to the smaller capacitance, almost as quickly as the short bitline
(the two curves are overlapped). On the other hand, precharging the bitline
after accessing a cell in the far segment (Figure~\ref{fig:substrate_sim_prech_on})
takes longer compared to the long-bitline baseline. As a result, \trp is reduced
for the near segment and increased for the far segment.

{\bf Summary (Latency, Power, and Die-Area).} Table~\ref{tbl:latency_comparison}
summarizes the latency, power, and die area characteristics of TL-DRAM compared
to short-bitline and long-bitline DRAMs, estimated using circuit-level SPICE
simulation~\cite{ibm_55nm} and power/area models from
Rambus~\cite{rambus-power}. Compared to commodity DRAM (long bitlines), which
incurs high latency (\trc) for all cells, TL-DRAM offers significantly reduced
latency (\trc) for cells in the near segment, while increasing the latency for
cells in the far segment due to the additional resistance of the isolation
transistor. In DRAM, a large fraction of the power is consumed by the bitlines.
Since the near segment in TL-DRAM has a lower capacitance, it also consumes less
power. On the other hand, accessing the far segment requires toggling the
isolation transistors, leading to increased power consumption. Mainly due to
additional isolation transistors, TL-DRAM increases die area by 3\% compared to
commodity DRAM. Section~4 of our HPCA 2013 paper~\cite{lee-hpca2013} includes
detailed circuit-level analyses of TL-DRAM, along with detailed area, latency,
and power estimations.

\begin{table}[ht]
\centering
\begin{footnotesize}
\setlength{\tabcolsep}{2.0pt}

\centering
\begin{tabular}{cccccc}

\toprule
& & Short Bitline & Long Bitline & \multicolumn{2}{c}{Segmented Bitline} \\
& &
(Figure~\ref{fig:intro_specialized_dram}) &
(Figure~\ref{fig:intro_commodity_dram}) &
\multicolumn{2}{c}{(Figure~\ref{fig:intro_tldram})} \\
\cmidrule{3-6}
& & Unsegmented & Unsegmented& Near & Far  \\
\cmidrule{1-6}

\multicolumn{2}{c}{Length (Cells)} & 32 & 512 & 32 & 480  \\

\cmidrule{1-6}

\multicolumn{2}{c}{Latency} & {\bf Low}  & High   & {\bf Low}  & Higher \\
\multicolumn{2}{c}{(\trc)}  & (23.1ns) & (52.5ns) & (23.1ns) & (65.8ns) \\

\cmidrule{1-6}

\multicolumn{2}{c}{Normalized}     & {\bf Low} & High   & {\bf Low} & Higher  \\
\multicolumn{2}{c}{Power} & (0.51) & (1.00) & (0.51) & (1.49)  \\

\cmidrule{1-6}

\multicolumn{2}{c}{Normalized}      & High    & {\bf Lower} & \multicolumn{2}{c}{{\bf Low}}      \\
\multicolumn{2}{c}{Die-Size (Cost)} & (3.76)  & (1.00) & \multicolumn{2}{c}{(1.03)}   \\

\bottomrule
\end{tabular}
\end{footnotesize}
\caption{Latency, power, and die area comparison.  Adapted from \cite{lee-hpca2013}.} \label{tbl:latency_comparison}
\end{table}

\section{Leveraging TL-DRAM} \label{sec:mechanism}

TL-DRAM enables the design of many new memory management policies that exploit
the asymmetric latency characteristics of the near and the far segments.
Section 5 of our HPCA 2013 paper~\cite{lee-hpca2013} describes four mechanisms
that take advantage of TL-DRAM. Here, we describe two approaches in particular.

In the first approach, the memory controller uses the near segment as a {\em
hardware-managed cache} for the far segment. In our HPCA 2013
paper~\cite{lee-hpca2013}, we discuss three policies for managing the near
segment cache. The three policies differ in deciding when a row in the far
segment is cached into the near segment and when the row is evicted. In
addition, we propose a new data transfer mechanism ({\em Inter-Segment Data
Transfer}) that efficiently migrates data between the segments by taking
advantage of the fact that the bitline is a bus connected to the cells in both
segments. By using this technique, the data from the source row can be
transferred to the destination row over the bitlines at very low latency
(additional 4ns over \trc).\footnote{A later work,
RowClone~\cite{seshadri-micro2013}, takes advantage of this property to enable
bulk copy and initialization completely within DRAM.} Furthermore, this
Inter-Segment Data Transfer happens exclusively within a DRAM bank without
utilizing the DRAM channel, allowing concurrent accesses to other banks.

In the second approach, the near segment capacity is exposed to the OS, enabling
the OS to use the full DRAM capacity. We propose two concrete mechanisms, one
where the memory controller uses an additional layer of indirection to map
frequently-accessed pages to the near segment, and another where the OS uses
static/dynamic profiling to directly map frequently-accessed pages to the near
segment. In both approaches, the accesses to pages that are mapped to the near
segment are served faster and with lower power than in conventional DRAM,
resulting in improved system performance and energy efficiency.

We refer the reader to Section~5 of our HPCA 2013 paper~\cite{lee-hpca2013} for a full description
of use cases for TL-DRAM. Note that a very wide variety of techniques developed
for cache management~\cite{seshadri-pact2012, qureshi-isca2006,
seshadri-taco2015, seznec-isca1993, tyson-micro1995} can be adopted to manage
the near segment in TL-DRAM.

\section{Performance and Power Evaluation}

Section~8 of our HPCA 2013 paper~\cite{lee-hpca2013} provides a detailed evaluation of all of
the above approaches to leverage TL-DRAM. Here, we present the
evaluation results for only the first approach, in which the near segment is used
as a hardware-managed cache managed under our best policy ({\em Benefit-Based
Caching}), to demonstrate the advantages of our TL-DRAM substrate. 

{\bf Methodology.} To evaluate our mechanism, we use 
Ramulator~\cite{kim-cal2015, ramulator}, an open-source DRAM simulator,
which is integrated into an in-house processor simulator. The released version of
Ramulator~\cite{ramulator} provides a model for TL-DRAM, which we hope future
works use and build upon.  A detailed methodology can be found in Section~7 of
our HPCA 2013 paper~\cite{lee-hpca2013}.

\begin{figure}[ht]
  \begin{subfigure}[b]{0.48\linewidth}
    \centering
    \includegraphics[width=\linewidth]{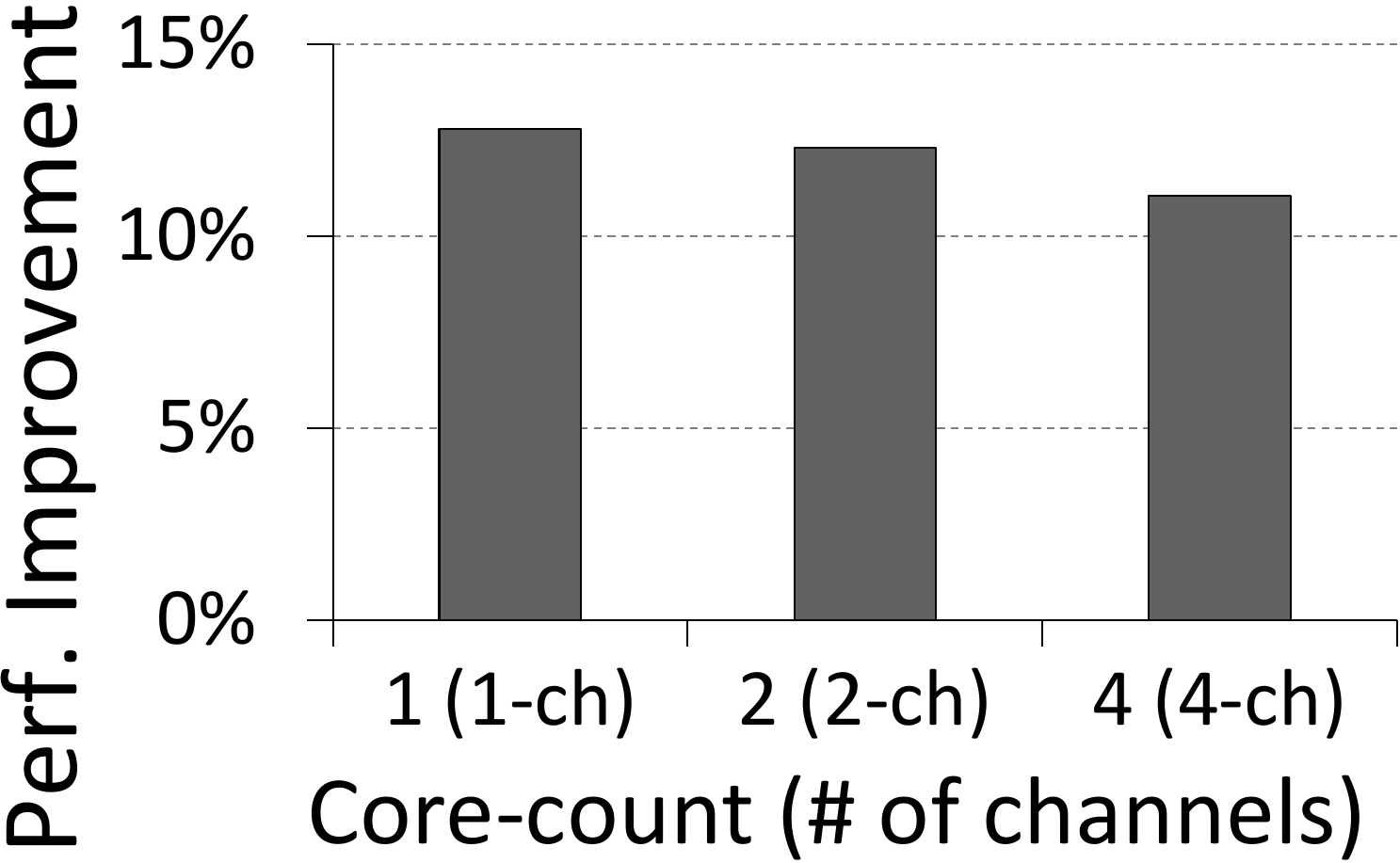}
    \subcaption{IPC improvement}
    \label{fig:result_ipc}
  \end{subfigure}
  \begin{subfigure}[b]{0.48\linewidth}
    \includegraphics[width=\linewidth]{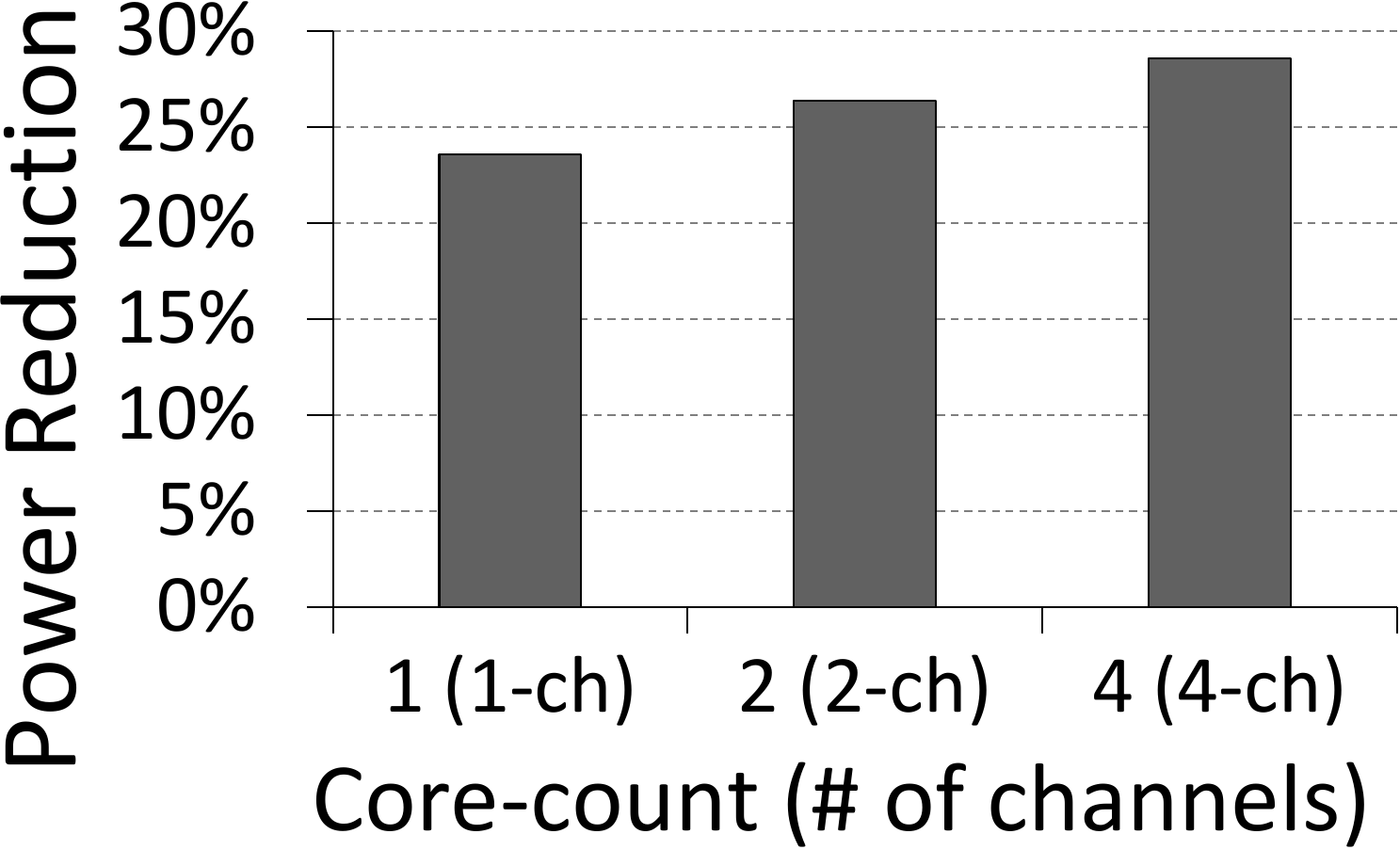}
    \caption{Power consumption}
    \label{fig:result_power}
  \end{subfigure}
	\caption{IPC improvement and power consumption of TL-DRAM. Adapted
	from~\cite{lee-hpca2013}.}
  \label{fig:result_cores}
\end{figure}

{\bf Performance \& Power Analysis.} Figure~\ref{fig:result_cores} shows the
average performance improvement and power efficiency of our proposed mechanism
over the baseline with conventional DRAM, on 1-, 2- and 4-core systems. As
described in Section~\ref{sec:tldram}, the access latency and power consumption
are significantly lower for near segment accesses, but higher for far segment
accesses, compared to accesses in a conventional DRAM. We observe that a large
fraction (over 90\% on average) of requests hit in the rows cached in the near
segment, thereby accessing the near segment with low latency and low power
consumption. As a result, TL-DRAM achieves significant performance improvements
of 12.8\%/12.3\%/11.0\%, and power savings of 23.6\%/26.4\%/28.6\% in
1-/2-/4-core systems, respectively.

\begin{figure}[ht]
  \includegraphics[width=\linewidth]{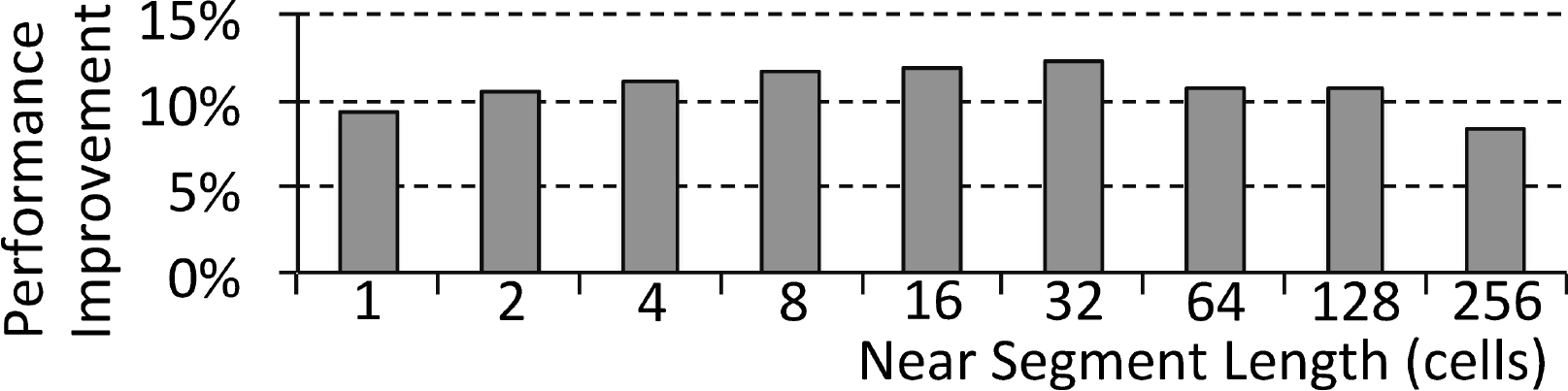}
  \caption{Effect of varying near segment capacity. Reproduced from~\cite{lee-hpca2013}.}
  \label{fig:result_single_sensitive}
\end{figure}

{\bf Sensitivity to Near Segment Capacity.} The number of rows in the near
segment presents a trade-off, since increasing the near segment's size increases
its capacity but also increases its access latency.
Figure~\ref{fig:result_single_sensitive} shows the performance improvement of
our proposed mechanisms over the baseline as we vary the near segment size.
Initially, performance improves as the number of rows in the near segment
increases, since more data can be cached. However, increasing the number of rows
in the near segment beyond 32 reduces the performance benefit due to the
increased capacitance and hence the higher near segment access latencies.

{\bf Other Results.} In our HPCA 2013 paper~\cite{lee-hpca2013}, we provide a detailed analysis of
how timing parameters and power consumption vary when varying the near segment
length (Sections~4 and 6.3 of \cite{lee-hpca2013}, respectively). We also
provide a comprehensive evaluation of the mechanisms we build on top of the
TL-DRAM substrate for both single- and multi-core systems (Section~8
of \cite{lee-hpca2013}).

	\section{Related Work}

To our knowledge, our HPCA 2013 paper~\cite{lee-hpca2013} is the first to {\em i)} enable latency
heterogeneity in DRAM without significantly increasing the DRAM cost per bit, and {\em
ii)} propose hardware/software mechanisms that leverage this latency
heterogeneity to improve system performance. We make the following major
contributions.

{\bf A Cost-Efficient Low-Latency DRAM.} Based on the key observation that long
internal wires (bitlines) are the dominant source of DRAM latency, our HPCA 2013 paper~\cite{lee-hpca2013} proposes a
new DRAM architecture called Tiered-Latency DRAM (TL-DRAM). To our knowledge
this is the first work to enable low-latency DRAM {\em without} significantly
increasing the DRAM cost per bit. By adding a single isolation transistor to each
bitline, we carve out a region within a DRAM chip, called the {\em near segment}, which
is fast and energy-efficient. This comes at a modest overhead of 3\% increase in
DRAM die-area. While there are two prior approaches to reduce DRAM latency
(using short bitlines~\cite{rldram, fcram}, adding an SRAM cache in
DRAM~\cite{hidaka-ieeemicro1990, hart-1994, esdram-2002, zhao-ieeemicro2001}),
both of these approaches significantly increase die-area due to additional sense
amplifiers or additional area for an SRAM cache, as we evaluate in our full
paper~\cite{lee-hpca2013}. Compared to these prior approaches, TL-DRAM is a much
more cost-effective architecture for achieving low latency.

There are many recent works that reduce {\em overall memory access latency} by
modifying DRAM, the DRAM-controller interface, and DRAM controllers. These works
enable more parallelism and bandwidth~\cite{kim-isca2012, chang-hpca2014,
seshadri-micro2013, lee-taco2016}, reduce refresh counts~\cite{liu-isca2012,
liu-isca2013, khan-sigmetrics2014, venkatesan-hpca2006, qureshi-dsn2015,
khan-cal2016, khan-micro2017, khan-dsn2016}, accelerate bulk
operations~\cite{seshadri-micro2013, seshadri-cal2015, seshadri-micro2015,
chang-hpca2016, seshadri-micro2017}, accelerate computation
in the logic layer of 3D-stacked DRAM~\cite{ahn-isca2015a, ahn-isca2015b,
zhang-hpca2014, guo-wondp2013, boroumand-cal2016, hsieh-iccd2016,
hsieh-isca2016, pattnaik-pact2016, liu-spaa2017, gao-hpca2016, kim.bmc18,
boroumand.asplos18}, enable better
communication between CPU and other devices through DRAM~\cite{lee-pact2015},
leverage process variation and temperature dependency in
DRAM~\cite{lee-hpca2015, chang-sigmetrics2016,
	chang-sigmetrics2017, chang-thesis2017, lee-thesis2016}, leverage design-induced variation in
DRAM~\cite{lee-sigmetrics2017}, leverage DRAM access
patterns~\cite{hassan-hpca2016, hassan-hpca2017, shin-hpca2014}, reduce write-related latencies
by better designing DRAM and DRAM control policies~\cite{chatterjee-hpca2012,
lee-techreport2010, seshadri-isca2014}, and reduce overall queuing latencies in
DRAM by better scheduling memory requests~\cite{mutlu-micro2007, mutlu-isca2008,
kim-hpca2010, kim-micro2010, subramanian-tpds2016, subramanian-iccd2014,
ipek-isca2008, usui-taco2016, ebrahimi-asplos2010, ebrahimi-isca2011,
ebrahimi-micro2011, moscibroda-usenix2007, muralidhara-micro2011,
lee-micro2008, ghose-isca2013, mukundan-hpca2012, kaseridis-micro2011,
hur-micro2004, shao-hpca2007}. Our proposal is orthogonal to all of these approaches and can
be applied in conjunction with them to achieve higher latency and energy
benefits.

{\bf Inter-Segment Data Transfer.} By implementing latency heterogeneity within
a DRAM subarray, TL-DRAM enables efficient data transfer between the fast and
slow segments by utilizing the bitlines as a wide bus. This mechanism takes
advantage of the fact that both the source and destination cells share the same
bitlines. Furthermore, this inter-segment migration happens only within a DRAM
bank and does not utilize the DRAM channel, thereby allowing concurrent accesses
to other banks over the channel. This inter-segment data transfer enables fast
and efficient movement of data within DRAM, which in turn enables efficient ways
of taking advantage of latency heterogeneity.

Other works that leverage latency heterogeneity in DRAM do not usually provide
any efficient mechanism of inter-segment data migration between different
latency segments. For example, Son et al.~\cite{son-isca2013} propose a low-latency DRAM
architecture that has different, fast (long bitline) and
slow (short bitline) subarrays in DRAM. This approach provides the significant
benefit only if latency-critical data is already allocated to the low-latency
regions (the low latency subarrays). Therefore, the overall memory system
performance is very sensitive to the page placement policy, and the system cannot
easily adopt to changes in the access latency of pages. In contrast, our new
inter-segment data transfer mechanism enables efficient relocation of pages,
leading to efficient dynamic page placement and relocation based on the
dynamically determined latency criticality of each page. Several more recent
works~\cite{chang-hpca2016, seshadri-micro2013, seshadri-cal2015,
seshadri-micro2017} take advantage of our concept of inter-segment data transfer
mechanism to perform page copy/initialization and bulk bitwise operations
completely within a DRAM chip.

\section{Potential Long-Term Impact}

{\bf Tolerating High DRAM Latency by Enabling New Layers in the Memory
Hierarchy.} Today, there is a large latency cliff between the on-chip last level
cache and off-chip DRAM, leading to a large performance fall-off when
applications start missing in the last level cache. By introducing an additional
fast layer (the near segment) within the DRAM itself, TL-DRAM smoothens this
latency cliff.

Note that many recent works add a DRAM cache or create heterogeneous main
memories~\cite{lee-isca2009, lee-ieeemicro2010, qureshi-isca2009,
meza-ieeecal2012, yoon-iccd2012, ramos-ics2011, satish-date2011, meza-weed2013,
luo-dsn2014, chatterjee-micro2012, ren-micro2015, li-cluster2017,
dhiman-dac2009, yu-micro2017} to smooth the latency cliff between the last level
cache and a longer-latency non-volatile main memory, e.g., \ch{phase-change
memory~\cite{lee-isca2009, qureshi-isca2009, qureshi-micro2009, meza-iccd2012,
yoon-taco2014, lee-cacm2010, lee-ieeemicro2010, wong.procieee10, meza-weed2013}},
STT-MRAM\ch{~\cite{kultursay-ispass2013, naeimi.itj13, meza-weed2013, wang-islped2014}}, or 
\ch{RRAM/memristors~\cite{wong-ieee2012, chua.tct71, strukov.nature08}}, or to take
advantage of the advantages of multiple different types of memories to optimize
for multiple metrics. Our approach is similar at the high-level (i.e., to reduce
the latency cliff at low cost by taking advantage of heterogeneity), yet we
introduce the new low-latency layer {\em within DRAM itself} instead of adding a
completely separate device. Tiered-Latency DRAM can also be used as a fast DRAM
cache.

{\bf Applicability to Future Memory Devices.} We show the benefits of TL-DRAM's
asymmetric latencies. Considering that most memory devices adopt a similar cell
organization (i.e., a two-dimensional cell array and row/column bus connections),
our approach of reducing the electrical load of connecting to a bus (bitline) to
achieve low access latency can be applicable to other memory devices.
Furthermore, the idea of performing inter-segment data transfer can also
potentially be applied to other memory devices, regardless of the memory
technology. For example, we believe it is promising to examine similar
approaches for emerging memory technologies like \ch{phase-change
memory~\cite{lee-isca2009, qureshi-isca2009, qureshi-micro2009, meza-iccd2012,
yoon-taco2014, lee-cacm2010, lee-ieeemicro2010, wong.procieee10, meza-weed2013}},
STT-MRAM\ch{~\cite{kultursay-ispass2013, naeimi.itj13, meza-weed2013, wang-islped2014}}, or
\ch{RRAM/memristors~\cite{wong-ieee2012, chua.tct71, strukov.nature08}}, 
as well as NAND flash memory
technology~\cite{luo.jsac16, cai-hpca2015, cai-dsn2015, cai-sigmetrics2014,
cai-iccd2013, cai-ieee2017, cai-hpca2017, cai-date2012, cai-date2013,
cai-iccd2012, luo-dsn2014, luo-msst2015, cai.procieee.arxiv17, cai.bookchapter.arxiv17,
luo.hpca18}.

{\bf New Research Opportunities.} The TL-DRAM substrate creates new
opportunities by enabling mechanisms that can leverage the latency heterogeneity
offered by the substrate. We briefly describe three directions, but we believe
that there are many new possibilities.

\begin{itemize}

	\item {\em New ways of leveraging TL-DRAM:} TL-DRAM is a substrate that can be
	utilized for many applications. Although we describe two major ways of
	leveraging TL-DRAM in our HPCA 2013 paper~\cite{lee-hpca2013}, we believe
	there are more ways to leverage the TL-DRAM substrate both in hardware and
	software. For instance, new mechanisms could be devised to detect data that is
	latency critical (e.g., data that causes many threads to become
	serialized~\cite{ebrahimi-micro2011, suleman-isca2010, joao-asplos2012,
	suleman-asplos2009, joao-isca2013} or data that belongs to threads that are
	more latency-sensitive or important~\cite{ausavarungnirun-pact2015,
	subramanian-hpca2013, subramanian-micro2015, ausavarungnirun-isca2012,
	subramanian-iccd2014, subramanian-tpds2016, kim-hpca2010, kim-micro2010,
	usui-taco2016, ebrahimi-asplos2010, lee-micro2008, lee-micro2009}) or could
	become latency critical in the near future and allocate/prefetch such data
	into the near segment.

	\item {\em Opening up new design spaces with multiple tiers:} TL-DRAM can be
	easily extended to have multiple latency tiers by adding more isolation
	transistors to the bitlines, providing more latency asymmetry. Our HPCA 2013
	paper~\cite{lee-hpca2013} provides an analysis of the latency of a TL-DRAM
	design with three tiers, showing the spread in latency for three tiers. This
	enables new mechanisms both in hardware and software that can allocate data
	appropriately to different tiers based on their access characteristics such as
	locality, criticality, priority, etc.

	\item {\em Inspiring new ways of architecting latency heterogeneity within
	DRAM:} To our knowledge, TL-DRAM is the first to enable latency heterogeneity
	within DRAM, which is significantly modifying the existing DRAM architecture.
	We believe that this could inspire research on other possible ways of
	architecting latency heterogeneity within DRAM~\cite{chang-sigmetrics2016,
	chang-sigmetrics2017, chang-thesis2017, lee-thesis2016, hassan-hpca2016,
	hassan-hpca2017, lee-sigmetrics2017, lee-hpca2015} or other memory devices. Note
	that recent works that are after our HPCA 2013 paper clearly exploit this
	promising direction proposed by our paper~\cite{chang-sigmetrics2016,
	chang-sigmetrics2017, chang-thesis2017, lee-thesis2016, hassan-hpca2016,
	hassan-hpca2017, lee-hpca2015, lee-sigmetrics2017,
	seshadri-micro2013}.

\end{itemize}

	\section*{Acknowledgments}
We thank Saugata Ghose for his dedicated effort in the preparation
of this article.
	Many thanks to Uksong Kang, Hak-soo Yu, Churoo Park,
Jung-Bae Lee, and Joo Sun Choi from Samsung, and Brian Hirano 
from Oracle, for their helpful comments. We thank the reviewers 
for their feedback. 
We acknowledge the support of our industrial
partners: AMD, HP Labs, IBM, Intel, Oracle, Qualcomm, and
Samsung. This research was also partially supported by grants
from the NSF (grants 0953246 and 1212962), GSRC, and
the Intel URO Memory Hierarchy Program.

	\bibliographystyle{sty/IEEEtranS}
	\bibliography{ref/ref_safari,ref/ref_others}

\end{document}